\ifdefined\relax \else
\pdfoutput=1
\fi
\documentclass[11pt,a4paper]{article}
\usepackage{iftex}
\ifPDFTeX
  \usepackage[T1]{fontenc}
  \usepackage[utf8]{inputenc}
\fi
\usepackage{lmodern}
\usepackage{amsmath,amssymb,amsthm,mathtools}
\usepackage[margin=25mm]{geometry}
\usepackage{microtype,booktabs,xcolor}
\usepackage[
  colorlinks=true,
  linkcolor=blue!40!black,
  citecolor=blue!40!black,
  urlcolor=blue!40!black,
  pdftitle={Rank Bounds for Half-Duplex Communication and a Linear Separation of Adversaries},
  pdfauthor={Timur Kuptsov}
]{hyperref}
\allowdisplaybreaks
\newtheorem{theorem}{Theorem}[section]
\newtheorem{lemma}[theorem]{Lemma}
\newtheorem{proposition}[theorem]{Proposition}
\newtheorem{corollary}[theorem]{Corollary}
\theoremstyle{definition}
\newtheorem{definition}[theorem]{Definition}
\newtheorem{remark}[theorem]{Remark}
\newcommand{\F}{\mathbb F_2}
\newcommand{\CC}{\mathrm{CC}}
\newcommand{\Ca}{C_{\mathrm a}^{\mathrm{hd}}}
\newcommand{\Ch}{C_{\mathrm h}^{\mathrm{hd}}}
\newcommand{\Cz}{C_{0}^{\mathrm{hd}}}
\newcommand{\Cs}{C_{\mathrm s}^{\mathrm{hd}}}
\newcommand{\EQ}{\mathrm{EQ}}
\newcommand{\DISJ}{\mathrm{DISJ}}
\newcommand{\GT}{\mathrm{GT}}
\newcommand{\IP}{\mathrm{IP}}
\newcommand{\rank}{\operatorname{rank}}
\title{Rank Bounds for Half-Duplex Communication\\and a Linear Separation of Adversaries}
\author{Timur Kuptsov\\\small Faculty of Mechanics and Mathematics, M.~V.~Lomonosov Moscow State University\\\small\texttt{timur.kuptcov@math.msu.ru}}
\date{November 2026}
\begin{document}
\maketitle
\begin{abstract}
We study deterministic half-duplex communication in which the participants may send or receive a bit in each round, with particular attention to the distinction between an honest adversary, which supplies the same bit to both receivers, and a malicious adversary, which may supply different bits. We prove a general rank lower bound in terms of a triangular fooling sequence and a real matrix whose zero pattern describes incompatible one-round events. In the malicious-adversary model the minimum possible rank is two. Consequently, the half-duplex complexity of equality, disjointness, and greater-than on $n$-bit inputs is exactly $n+1$. For the honest-adversary model we give a protocol with nonadaptive encoding rounds for equality of depth at most $n\log_3 2+\tfrac12\log_3 n+O(1)$, matching the known lower bound to first order. In particular, equality exhibits a linear separation between honest and malicious adversaries even though it is a total Boolean function. The two proofs illustrate complementary roles of algebraic incompatibility and combinatorial detection of unequal inputs.
\end{abstract}

\section{Introduction}
In deterministic half-duplex communication, Alice and Bob independently choose in each round whether to send a bit or receive. If exactly one sends, communication is classical; if both send, neither receives anything. The models differ in the outcome of a round in which both receive. An \emph{honest adversary} supplies a common arbitrary bit, whereas a \emph{malicious adversary} may supply different bits. The protocol must be correct for every sequence of allowed outcomes.

Hoover, Impagliazzo, Mihajlin, and Smal introduced half-duplex communication in connection with Karchmer--Wigderson games and formula complexity~\cite{HIMS18}. They also considered the \emph{silence model}, in which simultaneous receivers observe a distinguished symbol, and the \emph{zero model}, in which both receive zero. Writing $\Cs,\Cz,\Ch,\Ca$ for the four complexities and $\CC$ for classical deterministic communication complexity, with both parties required to output the answer, we have
\begin{equation}\label{eq:introchain}
\Cs(f)\le\Cz(f)\le\Ch(f)\le\Ca(f)\le\CC(f).
\end{equation}
For equality on $n$-bit strings, the silence and zero models have complexities
\[
\Cs(\EQ_n)=n\log_5 2+O(\log n),\qquad
\Cz(\EQ_n)=n\log_3 2+O(\log n),
\]
whereas the previously known malicious-adversary lower bound was $n/\log_2(5/2)$~\cite{HIMS18}. Vereshchagin and Dektiarev~\cite{DV25} separated malicious half-duplex communication from classical communication on a total non-Boolean function, with respective depths five and six. They also distinguished the honest model and gave a total non-Boolean function with honest complexity at most three and classical complexity four. Whether a total Boolean function admits a saving against a malicious adversary remains open.

We prove a general lower bound using a matrix of one-round event compatibilities. Let $N$ be a real matrix whose entries vanish exactly on incompatible event pairs. For a triangular fooling sequence of length $m$, any depth-$h$ protocol gives a triangular matrix with nonzero diagonal and rank at most $(\rank N)^h$. Consequently,
\[
h\ge\log_{\rank N}m.
\]
The nonzero entries of $N$ need not be positive or normalized. In the malicious model, this freedom reduces the rank from four for the zero-one compatibility matrix to two, the minimum possible value. We obtain, for every $n\ge1$,
\begin{equation}\label{eq:intro}
\Ca(\EQ_n)=\Ca(\DISJ_n)=\Ca(\GT_n)=n+1.
\end{equation}
The exact bound uses triangular sequences of length $2^n+1$: the additional cell is needed to obtain the final round beyond the usual $n$-round lower bound.

For an honest adversary, we construct a different equality protocol. A ternary constant-sum encoding guarantees a coordinate at which unequal inputs trigger a local discrepancy, including when both players receive. The resulting bound is
\begin{equation}\label{eq:introhon}
\Ch(\EQ_n)\le n\log_3 2+\tfrac12\log_3 n+O(1).
\end{equation}
Together with the zero-model lower bound, this determines the leading coefficient of honest equality and gives
\[
\Ca(\EQ_n)-\Ch(\EQ_n)=(1-\log_3 2)n+O(\log n).
\]
Thus honest and malicious adversaries have a linear gap on a total Boolean function. The algebraic lower bound and the encoding protocol are both proved for the same channel conventions; all lower bounds allow local outputs, and all constructed upper bounds have outputs determined by the terminal local history.

\section{Half-duplex protocols and compatible events}
Let $X,Y,Z$ be finite sets and let $f:D\to Z$ with nonempty $D\subseteq X\times Y$. A protocol has one rooted decision tree for Alice and one for Bob. At each vertex, Alice's action is determined by her input and her local history, and similarly for Bob. Each action is either $\mathrm{send}(0)$, $\mathrm{send}(1)$, or $\mathrm{receive}$. All branches can be padded to a common depth $h$.

If exactly one player sends, the other receives the transmitted bit. If both send, neither receives anything. If both receive, an honest adversary selects $c\in\{0,1\}$ and delivers $c$ to both; a malicious adversary selects an arbitrary ordered pair $(c_A,c_B)\in\{0,1\}^2$. We also refer to the \emph{zero model}, where both listeners receive $0$, and the \emph{silence model}, where they receive a distinguished silence symbol. Every possible outcome must be handled correctly. Each player must output $f(x,y)$ at the end. The smallest worst-case depth is denoted $\Ch(f)$ or $\Ca(f)$, respectively; $\Cz(f)$ and $\Cs(f)$ refer to the other two channels, and $\CC(f)$ is classical deterministic complexity.

The inequalities
\begin{equation}\label{eq:hier}
\Cs(f)\le\Cz(f)\le\Ch(f)\le\Ca(f)\le\CC(f)
\end{equation}
follow by simulating a more restrictive channel within a more permissive one or by using exclusively classical rounds. We use \emph{local-output protocols} for lower bounds: at a leaf, Alice's output may also depend on $x$ and Bob's on $y$. This only strengthens a lower bound. For total functions, local and global half-duplex complexity coincide in the adversarial models~\cite{DV25}.

Write $\mathsf s_i$ for sending $i$, and $\mathsf r_i$ for receiving $i$. A path in each tree is identified with its sequence of such events. In the malicious model the event alphabet is $E=\{\mathsf s_0,\mathsf s_1,\mathsf r_0,\mathsf r_1\}$. Two events, one from each player, are \emph{compatible} if they can occur during the same round. In order $\mathsf s_0,\mathsf s_1,\mathsf r_0,\mathsf r_1$ their compatibility matrix is
\begin{equation}\label{eq:compat}
K=\begin{pmatrix}1&1&1&0\\1&1&0&1\\1&0&1&1\\0&1&1&1\end{pmatrix}.
\end{equation}
Thus precisely $(\mathsf s_0,\mathsf r_1)$, $(\mathsf s_1,\mathsf r_0)$, and their transposes are impossible. In particular, two simultaneous receives are always compatible in the malicious model. In the honest model, by contrast, $(\mathsf r_0,\mathsf r_1)$ and $(\mathsf r_1,\mathsf r_0)$ are also impossible.

A path $\sigma$ is \emph{consistent} with an input $x$ if every sending event agrees with Alice's prescribed action at the relevant local vertex and every receiving event occurs when she chooses to listen. Two paths $\sigma,\tau$ are compatible if their events are compatible in every round.

\begin{lemma}[Realization of compatible paths]\label{lem:real}
If $\sigma$ is consistent with Alice's input $x$, $\tau$ is consistent with Bob's input $y$, and the paths are compatible, then a run of the protocol on $(x,y)$ follows exactly $(\sigma,\tau)$ under a suitable choice of channel outcomes.
\end{lemma}
\begin{proof}
Induct on the round number. Assume the players have reached the prescribed prefixes. Input consistency fixes each player's next send/receive action, including the transmitted bit when applicable. Compatibility guarantees that the requested pair of events is an allowed result of these actions. For simultaneous receiving, choose the stipulated allowed adversarial outcome; in all other cases the channel result is forced. Hence the prescribed next events occur. The induction reaches both specified leaves.
\end{proof}

\begin{corollary}[Crossed-run lemma]\label{cor:cross}
Suppose $(\sigma_i,\tau_i)$ is a run on $(x_i,y_i)\in D$, $i=1,2$. If $\sigma_1$ and $\tau_2$ are compatible and $(x_1,y_2)\in D$, then
\[
f(x_1,y_1)=f(x_1,y_2)=f(x_2,y_2).
\]
\end{corollary}
\begin{proof}
By Lemma~\ref{lem:real}, the crossed pair of paths is a run on $(x_1,y_2)$. Alice's local input and leaf are unchanged relative to the first original run, and Bob's relative to the second. Their respective outputs are therefore the two indicated original values; correctness of the crossed run forces both to equal $f(x_1,y_2)$.
\end{proof}

\section{A general rank lower bound}
The following ordered variant of a fooling set is adapted to the one-sided vanishing that produces a triangular matrix.

\begin{definition}\label{def:T}
A \emph{triangular fooling sequence} of length $m$ for $f:D\to Z$ is a sequence of input pairs $(x_1,y_1),\ldots,(x_m,y_m)\in D$ such that, whenever $i<j$, the crossed input $(x_i,y_j)$ belongs to $D$ and the three quantities
\[
f(x_i,y_i),\qquad f(x_i,y_j),\qquad f(x_j,y_j)
\]
are not all equal. Let $T(f)$ be the maximum possible length of such a sequence.
\end{definition}

Different diagonal values automatically satisfy the condition. If they agree, the upper-right crossed value must differ from them. No requirement is imposed on the lower-left crossed input. This directed choice is the reason the construction can yield better bounds than a method insisting on a symmetric zero pattern.

\begin{definition}\label{def:witness}
A \emph{rank witness} for a half-duplex model with event sets $E_A,E_B$ is a real matrix $N$, indexed by $E_A\times E_B$, satisfying
\[
N[e,e']=0\quad\Longleftrightarrow\quad(e,e')\text{ is incompatible}.
\]
All compatible entries must be nonzero, but may be negative. Write $r=\rank_{\mathbb R}N$.
\end{definition}

\begin{theorem}[Compatibility-rank bound]\label{thm:rank}
Every local-output protocol of depth $h$ computing $f$ in a model admitting a rank witness $N$ of rank $r\ge2$ satisfies
\begin{equation}\label{eq:rankbound}
h\ge \log_r T(f).
\end{equation}
\end{theorem}
\begin{proof}
Choose a triangular fooling sequence of length $m$ and fix one valid run $(\sigma_i,\tau_i)$ on each input $(x_i,y_i)$. Form the real $m\times m$ matrix
\[
Q_{ij}=\prod_{t=1}^{h}N[\sigma_i(t),\tau_j(t)].
\]
The matrix $Q$ is a submatrix of the Kronecker power $N^{\otimes h}$, allowing repeated rows or columns. Thus
\[
\rank Q\le\rank(N^{\otimes h})=(\rank N)^h=r^h.
\]
Each $Q_{ii}$ is nonzero because the chosen paths form an actual run, so all its event pairs are compatible. If $i<j$ and $Q_{ij}\ne0$, every event pair in $(\sigma_i,\tau_j)$ must be compatible. The crossed-run lemma would then imply
$f(x_i,y_i)=f(x_i,y_j)=f(x_j,y_j)$, contradicting the defining condition of the sequence. Hence $Q_{ij}=0$ for all $i<j$.

Consequently $Q$ is lower triangular with nonzero diagonal, and $\rank Q=m$. We conclude $m\le r^h$. Maximizing over the fooling sequence gives~\eqref{eq:rankbound}.
\end{proof}

The use of low-rank matrices with a prescribed zero pattern is related to the method of Haemers~\cite{Hae79}; the one-sided triangular formulation is related to Blokhuis-type arguments~\cite{Blo93}. This argument concerns the \emph{zero pattern}, not a probabilistic distribution of channel events. There is no requirement that the weights sum to one. In particular, cancellation between negative and positive weights is legitimate; imposing nonnegative or stochastic weights would discard the essential freedom.

\section{A rank-two witness for a malicious adversary}
The zero-one matrix~\eqref{eq:compat} has rank four: its determinant is $-3$. Nevertheless, the same support admits a rank-two matrix.

\begin{proposition}\label{prop:rank2}
The minimum real rank of a witness for the malicious-adversary model is exactly two. Thus every $f$ satisfies
\[
\Ca(f)\ge\log_2 T(f).
\]
\end{proposition}
\begin{proof}
Consider, in the event order fixed above,
\begin{equation}\label{eq:witness}
M=\begin{pmatrix}
1&1&1&0\\
1&-1&0&1\\
2&0&1&1\\
0&2&1&-1
\end{pmatrix}.
\end{equation}
Its only zero entries are $(1,4),(2,3),(3,2),(4,1)$, precisely the four incompatible pairs. The third row is the sum of the first two and the fourth is their difference; hence $\rank M\le2$. The first two rows are independent, so $\rank M=2$.

A rank-one witness is impossible: its first and fourth rows would have to be proportional whenever both are nonzero. But the first row has a zero in the fourth column, whereas the fourth row has a nonzero fourth entry; the fourth row has a zero in the first column, whereas the first row has a nonzero first entry. Therefore the minimum rank is two, and Theorem~\ref{thm:rank} applies with $r=2$.
\end{proof}

For an alternative factorization, set $e_0=(1,0)$, $e_1=(0,1)$, $h_0=(1,1)$ and $h_1=(1,-1)$. Assign to Alice's events the vectors $e_0,e_1,h_0,h_1$ and to Bob's events the vectors $h_0,h_1,e_0,e_1$. Matrix~\eqref{eq:witness} consists of their inner products. A depth-$h$ run therefore determines two tensors in $(\mathbb R^2)^{\otimes h}$ whose inner product is the corresponding entry of $Q$. The rank proof can be read as the impossibility of accommodating too many triangularly paired nonzero tensors in dimension $2^h$.

\section{Exact complexity of three Boolean functions}
For $x,y\in\{0,1\}^n$, define $\EQ_n(x,y)=[x=y]$ and $\GT_n(x,y)=[\operatorname{val}(x)>\operatorname{val}(y)]$, where $\operatorname{val}$ is the binary numerical value. Identifying strings with subsets of $[n]$, set $\DISJ_n(S,T)=[S\cap T=\varnothing]$. Here $[P]$ is one if $P$ holds and zero otherwise.

\begin{theorem}\label{thm:exact}
For every integer $n\ge1$,
\[
\Ca(\EQ_n)=\Ca(\DISJ_n)=\Ca(\GT_n)=n+1.
\]
\end{theorem}
\begin{proof}
The common upper bound is classical: Alice sends her $n$-bit input, after which Bob sends the one-bit answer. We construct a triangular fooling sequence of length $2^n+1$ for each function. Proposition~\ref{prop:rank2} then implies $2^h\ge 2^n+1$ and hence $h\ge n+1$.

For equality, take every diagonal pair $(x,x)$ once, in any order, followed by one off-diagonal pair $(a,b)$ with $a\ne b$. Two distinct diagonal pairs have value $1$, while their crossed input has value $0$. Every comparison with the final pair is automatic because its diagonal value is $0$.

For disjointness, identify $n$-bit strings with subsets of $[n]$. Use the pairs $(S,\overline S)$ for all $S\subseteq[n]$, ordered by nonincreasing cardinality, followed by $(\{1\},\{1\})$. The first $2^n$ pairs all have value $1$. If $S$ precedes a distinct $T$, then $|S|\ge|T|$ implies $S\not\subseteq T$, so $S\cap\overline T\ne\varnothing$ and the crossed value is $0$. The last pair has value $0$.

For greater-than, take the diagonal pairs $(x,x)$ in strictly decreasing numerical order, followed by $(1,0)$ (in the numerical interpretation of $n$-bit strings). Diagonal values are $0$; for $x>y$ the crossed value $\GT_n(x,y)=1$. The final pair has value $1$, completing the sequence.
\end{proof}

\begin{remark}
The malicious-adversary lower bound is stronger than the earlier estimates obtained from five-region round elimination~\cite{HIMS18,DISU21}. The leading improvement comes from the rank-two realization of the per-round compatibility pattern; the extra cell in the triangular sequence supplies the last round.
\end{remark}

\section{An equality protocol for the honest adversary}
The malicious-adversary exactness makes it natural to ask whether equality becomes easier when simultaneous listeners must receive the same bit. We now answer this affirmatively with an explicit family of protocols with nonadaptive encoding rounds.

A letter $p\in\{0,1,2\}$ specifies two roles, one for Alice and one for Bob, in a given round:
\[
\begin{tabular}{@{}cll@{}}\toprule
Letter & Alice's action & Bob's action\\\midrule
$0$ & receive, expect $0$ & send $0$\\
$1$ & send $0$ & send $1$\\
$2$ & send $1$ & receive, expect $1$\\\bottomrule
\end{tabular}
\]
Each player raises a persistent local alarm if, on a receiving round, the received bit differs from the expected bit. When both parties use the \emph{same} letter, there is no simultaneous receive and neither alarm is raised. For unequal letters, the following oriented condition is central.

\begin{lemma}[Letter test]\label{lem:letters}
If Alice uses letter $p$ and Bob uses letter $q$, then (i) $p=q$ raises no alarm, and (ii) $p<q$ raises at least one alarm under every honest-adversary choice, as well as in the zero model.
\end{lemma}
\begin{proof}
For $p=q=0$, Alice receives the expected $0$ from Bob; for $p=q=1$ both send; for $p=q=2$, Bob receives the expected $1$ from Alice. If $(p,q)=(0,1)$, Alice receives $1$ instead of the expected $0$. If $(p,q)=(1,2)$, Bob receives $0$ instead of $1$. If $(p,q)=(0,2)$, both receive. The honest adversary delivers a common bit $c$: Alice detects $c=1$, while Bob detects $c=0$. Thus an alarm occurs regardless of $c$. The zero model is the special case $c=0$.
\end{proof}

For $k\ge1$ let
\[
W_{k,s}=\{w\in\{0,1,2\}^k:\sum_{t=1}^k w_t=s\},\qquad 0\le s\le2k.
\]
Distinct words in the same $W_{k,s}$ are incomparable in the componentwise order. Indeed, if $u_t\ge v_t$ for every $t$ and the coordinate sums agree, then $u=v$. Thus if $u\ne v$, at least one coordinate satisfies $u_t<v_t$ (and at least one satisfies the reverse inequality).

\begin{theorem}[Honest equality protocol]\label{thm:honest}
Let $M_k=\max_{0\le s\le2k}|W_{k,s}|$. Equality on any set of at most $M_k$ inputs has an honest-adversary protocol, also valid in the zero model, of depth $k+2$. Consequently,
\[
\Ch(\EQ_n)\le n\log_3 2+O(\log n),
\qquad
\Ch(\EQ_n)=n\log_3 2+O(\log n).
\]
\end{theorem}
\begin{proof}
Choose an injective encoding $w:X\to W_{k,s}$ into a largest layer, with $|X|\le M_k$. In the first $k$ rounds, Alice performs her action for letter $w_t(x)$, and Bob performs his action for letter $w_t(y)$, without adapting subsequent actions to channel outcomes. Each accumulates a local alarm bit, $d_A$ or $d_B$.

If $x=y$, all letter pairs coincide, so Lemma~\ref{lem:letters} implies $d_A=d_B=0$. If $x\ne y$, the two words differ but have the same coordinate sum. Some coordinate satisfies $w_t(x)<w_t(y)$, and by the lemma at least one alarm is raised regardless of the adversary. Thus $d_A\lor d_B$ is exactly $[x\ne y]$.

Two final classical rounds make the result known to both parties. First Alice sends $d_A$ to Bob. Then Bob sends $\neg(d_A\lor d_B)$ to Alice. Both output the bit sent in this last round; it equals $[x=y]$.

The $3^k$ ternary words are partitioned into $2k+1$ layers. By the pigeonhole principle $M_k\ge3^k/(2k+1)$. Therefore a choice $k=n\log_3 2+O(\log n)$ makes $M_k\ge2^n$, proving the upper bound. The lower bound $\Ch(\EQ_n)\ge\Cz(\EQ_n)\ge n\log_3 2$ was proved for the zero model in~\cite{HIMS18}. Together they yield the stated asymptotic equality.
\end{proof}

\begin{proposition}[Second-order upper bound]\label{prop:central}
For $T_k=|W_{k,k}|$,
\[
T_k\sim\frac{3^{k+1/2}}{2\sqrt{\pi k}}.
\]
Moreover,
\[
\Ch(\EQ_n)\le\min\{k\ge1:T_k\ge2^n\}+2
=n\log_3 2+\tfrac12\log_3 n+O(1).
\]
\end{proposition}
\begin{proof}
Let $Z_1,\ldots,Z_k$ be independent, uniform on $\{-1,0,1\}$. Then $T_k=3^k\Pr(\sum Z_i=0)$. Fourier inversion gives
\[
\Pr\Bigl(\sum Z_i=0\Bigr)=\frac1{2\pi}\int_{-\pi}^{\pi}
\left(\frac{1+2\cos t}{3}\right)^k\,dt.
\]
Near zero the integrand base is $1-t^2/3+O(t^4)$. Outside any fixed neighborhood of zero its absolute value is bounded by a constant smaller than one. On a sufficiently small neighborhood it is bounded by $\exp(-ct^2)$ for some $c>0$. Substitution $t=u/\sqrt{k}$ and dominated convergence therefore give
\[
\Pr\Bigl(\sum Z_i=0\Bigr)\sim\frac1{2\pi\sqrt{k}}
\int_{\mathbb R}e^{-u^2/3}\,du=\frac{\sqrt3}{2\sqrt{\pi k}}.
\]
Apply the encoding protocol to $W_{k,k}$. The estimate $\log_3 T_k=k-\tfrac12\log_3 k+O(1)$ implies that the least admissible $k$ is $n\log_3 2+\tfrac12\log_3 n+O(1)$.
\end{proof}

\begin{corollary}[Linear separation]\label{cor:gap}
On total Boolean functions,
\[
\Ca(\EQ_n)-\Ch(\EQ_n)
=(1-\log_3 2)n+O(\log n),
\]
and the same asymptotic gap holds between $\CC(\EQ_n)$ and $\Ch(\EQ_n)$.
\end{corollary}
\begin{proof}
Theorem~\ref{thm:exact} gives $\Ca(\EQ_n)=n+1$, while the standard classical argument gives $\CC(\EQ_n)=n+1$. Apply Theorem~\ref{thm:honest}.
\end{proof}

\subsection*{Strictness of the model hierarchy}
The equality separation also completes the comparison of adjacent models on total functions.

\begin{proposition}\label{prop:hierarchy}
Each inequality in~\eqref{eq:hier} is strict for some total function.
\end{proposition}
\begin{proof}
The known silence and zero rates for equality separate the first two models~\cite{HIMS18}. Corollary~\ref{cor:gap} separates the honest and malicious models, and the five-versus-six example of~\cite{DV25} separates the malicious and classical models.

For the remaining comparison, let $\mathrm{OR}(x,y)=x\lor y$ on one-bit inputs. In the zero model a one-round protocol sends one on input one and receives on input zero. A sender outputs one; a receiver outputs the received bit. Thus $\Cz(\mathrm{OR})=1$.

Suppose an honest one-round protocol exists. A sender's view is independent of the other input, so Alice on input zero and Bob on input zero must receive. If Bob on input one sends $b$, Alice's view on $(0,1)$ equals her view on $(0,0)$ when the adversary supplies $b$. If Bob instead receives, the adversary can again make her views identical. In either case she cannot distinguish output zero from output one. Therefore $\Ch(\mathrm{OR})=2$, the upper bound being classical.
\end{proof}

\section{Limitations and open problems}
The triangular invariant can be small even when classical communication complexity is large. For inner product over $\F$, restrict a triangular fooling sequence to indices whose diagonal value is $c\in\{0,1\}$. The matrix
\[
A_{ij}=[x_i\cdot y_j=c]=x_i\cdot y_j+(1+c)
\]
is triangular with unit diagonal over $\F$. Its rank is at most $n$ when $c=1$ and at most $n+1$ when $c=0$. Hence $T(\IP_n)\le2n+1$, and Theorem~\ref{thm:rank} gives only a logarithmic bound for this function.

Three questions remain.
\begin{enumerate}\setlength{\itemsep}{.35em}
\item Does a total Boolean function satisfy $\Ca(f)<\CC(f)$? Theorem~\ref{thm:exact} rules out equality, disjointness, and greater-than as examples.
\item Is $\Ch(\EQ_n)=n\log_3 2+\tfrac12\log_3 n+O(1)$? Proposition~\ref{prop:central} gives the upper bound, while the available lower bound determines only the leading term.
\item Can a refinement of the compatibility-rank argument yield linear bounds for inner product or other functions with polynomially bounded triangular invariant? Such a refinement must use more than the single triangular matrix constructed here.
\end{enumerate}

\section*{Declaration on AI assistance}
Claude (Anthropic) and ChatGPT (OpenAI) assisted with research development and manuscript preparation. The author takes responsibility for the mathematical claims and states that he has proved and checked the results and proofs presented here.

\end{document}